# Integrating Neurophysiological Sensors and Driver Models for Safe and Performant Automated Vehicle Control in Mixed Traffic*

W. Damm, M. Fränzle, A. Lüdtke, J. W. Rieger, A. Trende, A. Unni

*Abstract*—In future mixed traffic Highly Automated Vehicles (HAV) will have to resolve interactions with human operated traffic. A particular problem for HAVs is detection of human states influencing safety critical decisions and driving behavior of humans. We demonstrate the value proposition of neuro-physiological sensors and driver models for optimizing performance of HAVs under safety constraints in mixed traffic applications.

## I. INTRODUCTION

Among the multiple challenges which must be addressed with the market introduction of highly autonomous vehicles (SAE levels 3 and higher) is the need to achieve high levels of safety while at the same time offering performance levels comparable to vehicles of lower SAE levels. These challenges include [1] "matching the perception capabilities of experienced human drivers under all environmental conditions within Operational Design Domain (ODD) such as recognizing all relevant objects within vehicle path and predicting future motions of all mobile objects (vehicles, pedestrians, bicyclists, animals…)". In the context of mixed-traffic applications, these challenges are further complicated by the need to understand and correctly interpret human-driver intent. If trajectories of Highly Automated Vehicles (HAVs) were determined by worst-case assumptions on human-driven vehicles, the resulting performance levels would be below thresholds acceptable by owners of HAVs. Even using empirically validated stochastic models of traffic participants, as e.g. advocated in [33], while significantly improving performance with respect to worst case behavior, do not allow further optimizations which are only possible when understanding driver-intent of those vehicles in the vicinity of the HAV critical for trajectory determination.

This paper proposes a unique combination of techniques stemming from different scientific communities to support on-line recognition of driver intent in vehicles in the proximity of the HAV ego vehicle. Specifically, our approach is based on recent results [2][3] on a particular class of neuro-physiological sensors, functional near-infrared spectroscopy (fNIRS) [4] where we demonstrated in driving simulator experiments the capability to differentiate driver states critical for maneuver selection based on measuring characteristics changes in oxygen saturation levels in particular brain regions discernable with fNIRS. We propose to exploit such cues about state changes by

- combining these with empirically validated driver models characterizing, in a given ODD and in a given driver state, the further moves of the driver, and by switching between such state-dependent driver models based on fNIRS detected changes of driver state;
- integrating such driver models in the HAV ego car and using Car2Car communication to forward queries about the driver-state of a particular vehicle in the HAV´s environment about the current driver state.

We demonstrate the feasibility of this approach in using recognition of driver state for gap selection in left and right-turns in intersections against oncoming traffic, where the ego HAV vehicle is approaching the intersection and uses a query to check the driver state awaiting a chance turning at the intersection. Our experiments show that the willingness to perform more risky maneuver increases with frustration level and/or sense of urgency. We observe characteristic changes in fNIRS measurements of certain brain areas when frustrated, corresponding to different levels of measured oxygen saturation. We show that the delay of the physiological response to changes of frustration levels of 6 seconds can be compensated by a combination of delay analysis in the actual dynamics of the induced left-turn strategy, and using control strategies coping with the resulting effective three second delay. We demonstrate the overall safety increase of the system utilizing such driver state detection, and analyze the performance gain stemming from intent detection, giving evidence to the industrial relevance of our approach.

This paper is structured as follows. Section 2 discusses related work. The main result is presented in Section 3, with subsections describing the overall system decomposition, the experimental setting, the neuro-physiological measurements, the driver model, the control strategy, the safety impact and the overall performance in Section 4. The conclusion outlines future work.

## II. RELATED WORK

Operating a vehicle is a complex safety-critical task since different cognitive demands are concurrently imposed on the driver because information from traffic signs, in-vehicle

*Research supported by the German Science Foundation under Grants DA 206/11-1, FR 2715/4-1, LU 1880/2-1 and RI 1511/2-1

W. Damm (phone +49-441-9722-500; e-mail: werner.damm@uni-oldenburg.de) and M. Fränzle (phone +49-441-9722-566, e-mail: martin.fraenzle@informatik.uni-oldenburg.de) are with the Department of Computer Science, University of Oldenburg, 26121 Oldenburg, Germany.

A. Lüdtke (phone +49-441-9722-530; e-mail: luedtke@offis.de) and A. Trende (phone +49-441-9722-231; e-mail: alexander.trende@offis.de) are with OFFIS, 26121 Oldenburg..

J. W. Rieger (phone +49-441-798-4533; e-mail: jochem.rieger@uol.de) and Anirudh Unni (phone +49-441-798-5167; e-mail: anirudh.unni@uol.de) are with the Department of Psychology, University of Oldenburg, 26129 Oldenburg, Germany.

displays, and other traffic participants has to be integrated into a coherent situation representation. Driving task performance is strongly influenced by cognitive and emotional processes of the driver. Therefore, being able to reliably measure the momentary cognitive or emotional state of the driver would be a major step into the direction of designing automation systems that are adaptive to the driver's state.

### A. Peripheral vs central physiological measurement

Activation in the autonomous nervous system is modulated by various cognitive and emotional factors, and peripheral physiological measures such a pupil dilation, heart rate, and blood pressure are used as overt, indirect measures to gauge workload level effects on the autonomous system. Several studies attempted to assess changes in cognitive workload levels in realistic situations from peripheral physiological parameters such as heart rate, heart rate variability or skin conductance level [5][6]. However, relying on peripheral physiology has the disadvantage that changes in arousal are not specific to cognitive states, but are also integral to emotional states such as anger [7] or/and related to physical activity or fatigue [8].

Another method for measuring cognitive and emotional states is via direct measurement of mental resources, the fuel that allows for cognitive processing [9]. Brain activations appear to be the most natural and direct measure because the brain activation is the immediate physiological basis of mental work [10], emotions, appraisal processes, subjective experiences [11] and the specificity of these subjective state is paralleled in specific multidimensional spatiotemporal brain activation patterns. By exploiting these links between subjective states and brain activation, neuroimaging allows for objective non-invasive measurement of mental resources during variable task load. Measuring multidimensional brain function offers some unique advantages. Continuous measurements can extract covert subjective states continuously in complex environments in which overt responses by the subject may be relatively sparse [12]. The high dimensionality of the brain data is a necessary prerequisite for discrimination of multiple simultaneously changing cognitive and emotional states and characterization of their interactions to provide reliable data to discriminate and quantify different driver states with relatively high accuracy.

In our previous work, we demonstrated the feasibility of temporally continuous prediction of variations of cognitive working memory load (WML) over multiple levels in a realistic driving scenario with multiple parallel tasks [2]. Therefore we used a predictive modelling approach based statistical learning to exploit the increased spatial specificity of high-density whole head fNIRS. FNIRS allows for spatially resolved measurement of brain activation related changes in blood oxygenation levels brain with good spatial and acceptable temporal resolution. In addition to time resolved workload assessment the analysis of the learned predictive model suggested a network of bilateral inferior frontal and bilateral temporo-occipital areas as being specifically involved in working memory load related processing.

### B. Modelling of human driving behavior with cognitive architectures

Human driving behavior with respect to the cognitive processes involved has been studied in the last decade by numerous researchers. [29] proposed one of the first driver models implemented in ACT-R. Similar approaches have been used in lane-merging models in [30][31]. [32] used fNIRS in combination with a driving simulator to investigate the subject's working memory load. Using an n-back task the workload was manipulated. The neurophysiological workload measures were compared to the workload predictions of a virtual driver model implemented in the cognitive architecture CASCaS. The authors hypnotize that an adaptive driver assistance system based on human modelling could benefit from the input of the driver's current workload level based on neurophysiological measurements.

### C. Decision making models in traffic

The modelling of human driving can have important implications for road capacity and safety. Models of human driving behavior in lane-change and turning situations have been proposed among others in [21][22][23]. [24] presents a decision model for gap-acceptance for unsignalized intersections. It is argued that the threshold for gap acceptance, called critical gap is not the same for every traffic participant. Factors that contribute beside others are the age, gender and driving style of the driver. [25] came to a similar conclusion when they developed a passing gap acceptance model for highway situations. They found that not just gap attributes, like gap size relative velocities, but also personality characteristics like, gender or age, affect passing behavior significantly.

## III. TECHNICAL REALIZATION

Turning to the development of the intended safety function, which tightly integrates neurophysiological measurements, driver intent recognition, and autonomous control, we follow a system engineering approach sketched underneath. It starts from an appropriate system decomposition and then elaborates on the –factually geometrically distributed- system components individually.

### A. System Decomposition

The intended functionality of our system is as follows: An HAV vehicle approaching an intersection and willing to traverse it in a straight line (cf. Fig. 4) shall be observant to oncoming traffic waiting to pursue turning. Depending on the expected gap acceptance strategy of the human driver in the first car waiting for turning, the HAV shall select and pursue risk-mitigating driving actions, namely to either actively widen or close the gap in front of it and thereby adapt to and disambiguate the situation. This in turn requires reliable detection of the gap acceptance strategy of the particular human driver in the particular situation. To realize such functionality, the function architecture demands the following sub-functions, which are subsequently to be refined and allocated.

1. Beyond standard environmental sensing in the HAV for detecting objects in the vicinity, determining absolute and relative speed, gap size, traffic lights and signs, etc., the system needs *sensors for determining situational and individual variations in human user state*, as relevant to the driving strategy.

   In our case, the crucial human state is the frustration level of the human driver in the car waiting for a turn. Such frustration develops over time, with the slope of development depending on external factors like felt or actual time pressure, and is the cause for a gradual shift to more risky gap acceptance strategies.

   Unfortunately, neither frustration level nor perceived time pressure are directly observable or measurable by technical sensors, hence we have to employ measurement of neurophysiological correlates instead. Oxygen saturation in certain cortical regions turns out to be strongly correlated, exhibiting a characteristic slope leading to a reliably detectable increase in regional fNIRS (functional near-infrared spectroscopy, see Section III.C) measurements. Owing to the dynamics of the underlying metabolic processes, such measurements come with a phase-delay of approximately 6s in relation to the human state they are indicative for. This phase delay will have to be accounted for by the function layout.

   Furthermore, fNIRS measurements require contact to the subject such that they can only be taken by equipment in the human-driven car, enforcing a distributed system design allocating some components in the HAV and some sensor equipment in the human-driven car, to be connected by car2car communication and forming a networked control system.

2. In order to determine the gap acceptance strategy to be expected from the particular human driver in the particular situation, the HAV needs a *valid model of the human's gap acceptance strategy* in relation to observable or known parameters like driver's gender and frustration level, with the latter being determined by the aforementioned measurement modality. The model needs to be corrected for the human subject's tendency to cooperate differently with an HAV than with a human-driven alter car. The model (to be developed in Section III.D) has to be executable and will become part of the control functionality embedded into the HAV, where it constitutes a part of the HAV's tactical control system's world model (see Section III.E).

   As the HAV operates in a highly safety-critical context, the embedded model of human behavior needs a thorough experimental and statistical validation, which is the subject of the section III.D.

3. As gap acceptance varies depending on whether the human driver classifies the oncoming car as an HAV or as manually driven, the HAV needs a clearly visible external marker identifying it as an HAV. We will not elaborate on design of such markers, but assume them given.

4. The HAV needs a *strategy at tactical control level that maps* the observed (historic and current) situation into an adaption of its own driving strategy. Tactical control hereby has to decide whether to widen or narrow the gap in front of it. As the controller has to resort to a perception of the situation which is subject to considerable latencies especially in sensing the human state, the control strategy has to actively compensate for these delays, as shown in Section III.E).

5. The HAV finally applies *low-level control* to implement the tactical decisions. Such low-level control implements control skills like adjusting relative distance and speed, as necessary to operationalize the tactical decisions concerning change of gap size. As such control skills are standard algorithms of continuous control, we do not further elaborate on them in this paper.

*B. Neurophysiological sensing*

FNIRS is a non-invasive optical neuroimaging technique to measure hemodynamic responses in the brain [13]. FNIRS uses low energy optical radiation in the near-infrared range (wavelength 700-900 nm) to measure absorption changes that reflect local concentration changes of oxygenated hemoglobin (HbO) and deoxygenated hemoglobin (HbR) in cortical brain areas. The near-infrared light spectrum takes advantage of the fact that it falls in the optimal wavelength window in which skin, tissue, and bones are relatively transparent to the electromagnetic spectrum with little absorption and mostly scattering, while HbO and HbR are the stronger absorbers of light. The differences in absorption spectra of HbO and HbR allow us to measure relative hemoglobin concentration changes through the use of light attenuation at multiple wavelengths.

FNIRS relies on the principle of neurovascular coupling also known as the hemodynamic response or blood-oxygen-level dependent (BOLD) response [13][14]. Although local variations of the blood oxygen level are caused by temporally varying neural activity, the hemodynamic response lags the neuronal activation change by 4-6 s. Figure 1 shows a schematic hemodynamic response to a brief neural activation. In the next section, we present a case study where we investigated if cortical markers for frustration while driving could be possible based on whole-head fNIRS brain activation measurements.

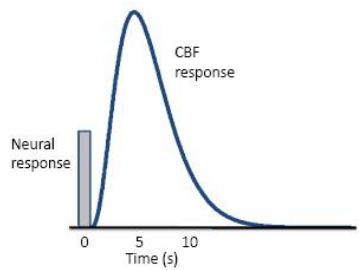

Figure 1. Schematic of hemodynamic response to brief neural activation[1]

*C. Case study: Recognizing driver frustration*

Experiencing frustration while driving can harm cognitive processing, result in aggressive behaviour, and hence negatively influence driving performance and traffic safety. Being able to automatically detect frustration would allow adaptive driver assistance and automation systems to adequately react to a driver's frustration and mitigate potential negative consequences. To identify reliable and valid indicators of driver's frustration, we measured cortical activation from almost whole-head fNIRS while participants experienced six frustrating (Frust) and non-frustrating (noFrust) driving simulator scenarios [3]. We induced frustration through a combination of time pressure and road-blocking elements and applied machine learning methods to predict the subjective frustration levels from brain activation. We performed univariate generalized linear model (GLM) analyses separately for each channel in order to determine the localization of brain areas most predictive to frustration while driving. This revealed relative concentration increases of HbO and decreases of HbR during Frust drives compared to NoFrust drives in brain areas known to be involved in cognitive appraisal, impulse control and emotion regulation processes. Figure 2 depicts the block averaging of the fNIRS time-series HbO and HbR data from the six Frust and noFrust scenarios for an example channel from the left prefrontal cortex for an example subject.

---
[1] http://www.scholarpedia.org/article/Neurovascular_coupling

Frustrated driving intervals were indicated by increased fNIRS brain activation in the bilateral inferior frontal, bilateral ventral motor, and left temporo-occipital cortices. Figure 3 shows the results presented as unthresholded t-statistic maps (difference: Frust-noFrust) from the channel-wise linear regression of HbR and HbO data for the group level analysis. The t-statistic maps indicate the local effect sizes, in essence they are Cohen's d scaled by the square root of the number of samples included in their calculation. The t-values provide a univariate measure to estimate the importance of a feature for multivariate classification.

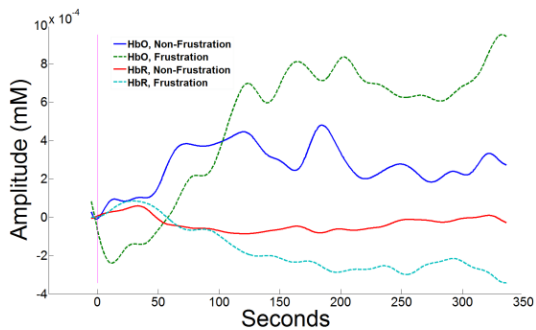

Figure 2. Block averaging of time-series data for an example channel in the left prefrontal cortex for an example subject.

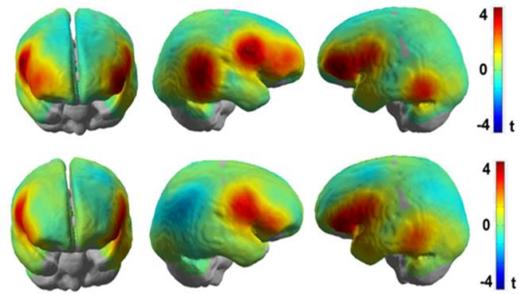

Figure 3. t-statistic maps obtained from group-averaging channel-wise linear regression of (a) HbR and (b) HbO fNIRS data over the Frust and noFrust conditions using Generalized Linear Model. High positive t-values are indicated by red colour, high negative values by blue colour. Figure taken from Ihme et al. 2018.

As a next step, a multivariate logistic ridge regression [15] decoding model was implemented in the Glmnet toolbox [16] for the prediction of Frust and NoFrust drives from sample-by-sample fNIRS brain activation data. The multivariate approach combined brain measurements spatially over different sensors to get a model output of the logistic regression model that can be interpreted as a class probability. The input features that went into the decoding model were the pre-processed HbO and HbR values which were z-scored for the particular segments of Frust and noFrust drives. The model weighted these input features and provided an output between 0 and 1. This output value indicated the likelihood for the test data classified as either the Frust class or the noFrust class. Using this approach allowed us to discriminate between frustrated and non-frustrated driving intervals with a relatively high accuracy of 78.1 % (mean over 12 participants).

Our results showed that cortical markers of frustration can be informative for time resolved driver state identification in complex realistic driving situations [3].

*D. Human Model*

Waiting at an intersection while feeling a sense of urgency can potentially cause frustration [18]. To investigate the gap acceptance in turning situations under time-pressure, we performed a driving simulator study in a full-scale fixed-base driving simulator. We hypothesized that the act of waiting for a gap at an intersection can cause frustration when the subject fells a sense of urgency. The study was conducted with 17 subjects (7 males, 10 females, mean age = 26.0y, SD age = 9.3y, mean driving experience = 8.6y) and has been described in more detail in [17].

Each subject encountered multiple intersections during one experimental block. At each intersection, the subjects had to stop because of a stop sign and the arriving traffic (cf. Fig. 4). Except for one wider gap, the cars in the arriving traffic kept a time headway that made merging impossible. The first car after the gap could be either an HAV or a normal human-driven car. At the intersection, the subjects could choose to either merge into the gap or wait until all the traffic passed hence losing time in the block. HAVs were easily distinguishable from other cars by car model and color. Additionally, they did not contain a virtual human model inside the car. The subjects were told beforehand that the HAV were defensively programmed to avoid collisions. In reality all cars followed the same driving behavior during the simulation, which the participants did not know. The experiment consisted of three blocks: the first block was a training session followed by one session with and one without a time limit. If the subjects managed to finish the time limit block in the given time limit they got an additional monetary reward. It was necessary for the subjects to take some of the gaps instead of just waiting for the traffic to pass to reach the end of the scenario within the given time limit.

We use a Bayesian Network to model the merging decision in the described experiment. A Bayesian Network is a probabilistic graphical model and directed acyclic graph. Nodes represent the variables of the model. Their conditional dependencies are represented as edges in the graph [19].

We have chosen a rather abstract model considering just four binary random variables: "Merging" (M), „Time Limit" (TL), „Interaction partner" (IP) and „Gender" (G) (cf. Fig 5). As described above, these will not be the only factors describing the probability of a merging decision. TL, IP and G are independent of each other and each one is a parent node of M. The corresponding joint probability function is:

$$p(M, TL, IP, G) = p(M|TL, IP, G)\ p(TL)\ p(G)\ p(IP), \quad (1)$$

where $M \in \{true, false\}$, $TL \in \{true, false\}$, $G \in \{M, F\}$, and $IP \in \{AV, H\}$.

We set the probabilities for the parent nodes according to the experimental setting. It will be argued later that these numbers will most likely not represent real traffic situations. Every subject drove one block with and one without a time limit which leads to $p(TL = true) = 0.50$. In the study described above, we had twice as many female subjects. Some data had to be excluded from the data analysis leading to $p(G = F) = 0.59$ instead of $p(G = F) = 0.66$. The exclusion of this data did not change the ratio between interactions with AV and humans significantly. Thus $p(IP = AV) = 0.50$.

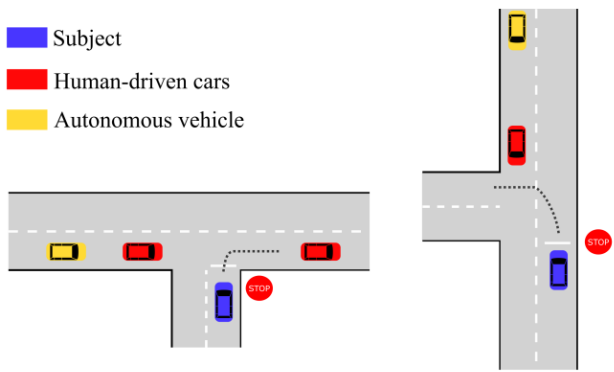

Figure 4. Sketch of the intersections used in the driving simulator study: The subject is waiting at an intersection. The arriving traffic (red) includes an HAV (yellow). The subject has the option to merge into the gap or wait for the traffic to pass and turn.

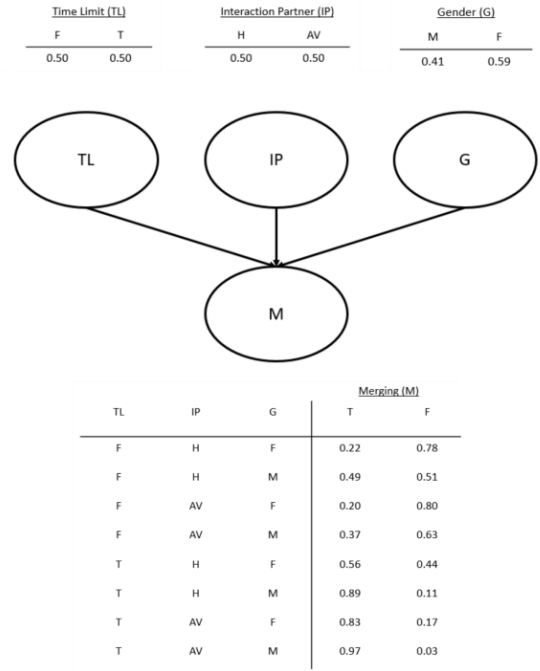

Figure 5. Bayesian network structure for the human merging model. The three parent nodes "Time Limit", "Interaction Partner" and "Gender" each are connected to the "Merging" node. The corresponding conditional probability tables are shown next to each node.

One advantage of modelling the merging decision with a Bayesian Network is their capability for probabilistic reasoning under uncertainty. Let us assume for example that the HAV wants to infer the merging decision of a waiting car at the intersection, but the evidence for the Time Limit cannot be set. This could be due to the car having no neurophysiological sensor installed or the recognition of the neurophysiological state is not reliable in the given case. In this case, the model could still make assumptions about the merging probabilities.

To demonstrate this we compare four different models. Each model differs in the number of observed parent variables. We calculated the Receiver Operating Characteristic (ROC) for a 10-fold cross-validation and compare the mean accuracy, false-negative-rates (FNR) and mean area-under-curve (AUC) for 10 repetitions of cross-validation. The FNR corresponds to the probability that the subject merged but the model predicts that the subject waited. These are the cases that can lead to accidents. The results are shown in Table 1, whereas $p(M | TL, IP, G)$ means that evidence for all three parent variables was used. Overall, the subjects merged in 55% of the intersections.

The addition of evidence leads to overall better values for AUC and accuracy. The knowledge about whether the subject was acting under time pressure helps to predict his/her decision. Although the accuracies and AUCs between the three models including Time Limit don't differ much, the FNR is lower for the $p(M | TL, IP)$ model and should be preferred to minimize the probability of accidents.

We have chosen the probabilities of the parent nodes, such as Gender, according to the experimental setting. [20] found though, that in 2010 50.3% of the US drivers were female.

However, male drivers drive 59% of the miles driven on the road. Therefore, it is more likely to encounter a male driver in traffic. According to this statistics it would be more reasonable to choose p(G = F) = 0.41 and p(G = M) = 0.59. Additionally, the probability to encounter HAV on the street has to be adjusted to current use in traffic. It has to be mentioned that it was clear for our subjects how to differentiate between autonomous and human driven cars. This difference may not be obvious for all drivers in real traffic. Some subjects also mentioned in an interview after the experiment that the fact that no human was inside the HAV was relevant for their merging decisions. The Interaction Partner variable should therefore be extended and have a state corresponding to an HAV with one or more passengers inside. In real traffic situations, the number of drivers who experience urgency during their ride is not clear. The data collected by a broad usage of the above proposed neurophysiological sensor could give an approximation for this variable. Factors like time-to-collision, waiting time and other, as investigated in studies like [18][28] should also be included to create a more accurate model.

TABLE I. MODEL RESULTS

|  | p(M \| TL, IP, G) | p(M \| IP, G) | p(M \| TL, G) | p(M \| TL, IP) |
|---|---|---|---|---|
| AUC +- SD | **0.81 +- 0.01** | 0.62 +- 0.01 | 0.78 +- 0.07 | 0.80 +- 0.01 |
| Accuracy +- SD | **0.80 +- 0.01** | 0.68 +- 0.01 | 0.78 +- 0.01 | 0.78 +- 0.01 |
| FNR +- SD | 0.26 +- 0.02 | 0.30 +- 0.04 | 0.27 +- 0.03 | **0.22 +- 0.03** |

Table 1: Results of the model selection with 10-fold cross validation. We calculated the area under curve (AUC), accuracy and false-negative rate (FNR) for four different models over 10 repetitions of cross validation. Models that include evidence about the Time Limit lead to overall better results w.r.t to mean AUC, accuracy and FNR.

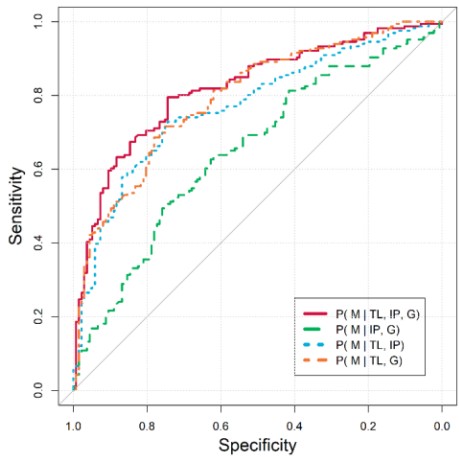

Figure 6. Example Receiver Operating Characteristic for one iteration of the 10-fold cross validation and all four models. Different evidences from the data was used for the prediction of a merging decision. Models including evidence about the Time Limit have overall higher AUC values.

Right now, the feeling of urgency is just incorporated via the node Time Limit thus representing the presence of a time constraint for the subject. As shown and discussed in section III.C. another node representing the classifier result of a neurophysiological frustration measure could be included into the Bayesian Network. We hypothesize that this node would not be independent of every parent node so far and therefore creating a more complex probability function represented by the graph structure.

*E. Control Synthesis*

The tactical control layer can be synthesized automatically after formulating it as a 2½-player game over hybrid discrete-continuous state and of imperfect information, where

- the two players are the HAV and the human-operated vehicle waiting for a turn,
- the half player encodes uncontrollable stochastic influences,
- the moves in the game encode the control actions available to the two players (including the "spontaneous", not further determined "decision" of the human to feel stressed, which shows 6s later by being mirrored in an information item observable by the HAV),
- the winning condition encodes avoidance of close encounters between HAV and human-operated oncoming traffic.

The synthesis objective then is to construct a strategy for the HAV which ensures, for any choice of actions by the human driver, that the probability of a win for the HAV is above a societally agreed threshold. As a win corresponds to avoidance of a close encounter, this is equivalent to searching for a strategy guaranteeing that the likelihood for close encounters remains below a societally accepted threshold in any possible scenario, where the space of possible scenarios is spanned by the moves of the human and the according reactions of the HAV.

As the temporal horizon of this 2½-player game is bounded due to the finite duration of the crossing situation, solvers like SiSAT [26] can solve such games. The game representation encodes the various delays associated with the control problem and thus forces the strategy synthesis to construct a tactical control strategy that is resilient to these delays. To this end it is worthwhile noting that some delays are detrimental to controllability, like the delayed observation of human state induced by the metabolic process registered by fNIRS measurements, while others are advantageous, like the time it takes for the waiting human-operated car to actually start from stand-still and progress into opposing traffic. With the former delay being approximately 6s and the latter in the range of 2s, the controller has to make sure that it draws decisions well ahead of time which are robust enough to cover the residual ca. 4s latency. Controller synthesis algorithms can do so systematically. [27] expose a practical algorithm combining feasible computational costs with completeness, i.e., finding a delay-resilient winning strategy whenever such exists, and then apply that algorithm to numerous collision avoidance problems in competitive and geometrically confined situations, thereby demonstrating feasibility of safe control under delayed situation awareness in collision-avoidance situations.

The controller analyzed underneath (Section III.F) for its safety impact has partially been obtained by such synthesis.

As the game model currently remains incomplete due to only partial coverage of possible situations in the experimental setup, it had to be refined manually. Its operational principle can be paraphrased as follows:

1. When approaching the intersection, the HAV already from a distance of 6 gaps ahead starts to continuously check whether the fNIRS equipment in the first manually driven car waiting for a left turn reports frustration. It combines all these reports by logical OR, making sure that even a transient positive detection becomes persistent. The OR-ed together signal is only cleared when the HAV detects that the particular car has taken a gap such that a new car is now waiting for a left turn (or none at all). In this case, signal collection starts fresh.
2. When there is positive indication of frustration, the HAV immediately opens the gap in front of it by decelerating, unless the gap has already been wide enough for safe passage of the manually operated alter car. Without detected signs of frustration, it just maintains the distance suggested by its general driving strategy, which likely is too tight to permit safe passage.

*F. Analysis of Safety Impact*

We now demonstrate the overall safety of the system and analyze the performance gain stemming from intent detection, giving evidence to the industrial relevance of our approach.

As evident from the conditional probabilities stated in Fig. 5, humans show a considerably stronger situational variation in behavior when interacting with an HAV than when interacting with a human-operated car. While in absence of time pressure, they are substantially more likely to grant preference to the HAV than to the humans, this relation reverses when under time pressure. This indicates a strong belief in politeness and situational adaption of autonomous driving functions, which would however be unjustified without sensing technologies for human state and state-dependent driving strategies as suggested in this article. A positive safety impact of the technology thus seems probable.

When quantifying that impact, we need the following figures:

1. The sensitivity and specificity of the fNIRS-based detection of frustration. A conservative, safety-oriented design would adjust the detector for a moderately high sensitivity even if that may come at the price of specificity, as falsely positive verdicts ("not frustrated" when actually frustrated) are a prerequisite for the human likely engaging into risky maneuvers without providing awareness of this situation to the HAV. For the sake of an example, let us assume that we calibrated the detection system for frustration to a sensitivity of 0.78 in accordance with the findings of Section III.B.
2. We need to have a model of the build-up of frustration. For the sake of demonstrating the shape of a quantified safety case, we have derived an initial model obviously requiring further experimental validation directly from the slope of frustration visible in Figure 2. The model used as an example assumes frustration to build up over the number of gaps that passed without permitting a passage, whereby the average number of gaps leading to frustration was set to 8 with a variance of 2 gaps within the formal analysis in order to confine computational complexity.

Using these values, we generated a corresponding symbolic representation of a hybrid-state Markov decision process (MDP) in SiSAT syntax. Within this MDP, gap sizes in traffic on the HAV's lane as well as occurrence times of manually driven cars in opposing traffic were existential variables, forcing SiSAT to construct a worst-case (i.e., maximally risky) scenario, while gap acceptance, build-up slopes for frustration, as well as the frustration detection were random variables. SiSAT was thus asked to construct a worst-case scenario of short and long gaps leading to maximum risk. The probabilities for the random variables were directly taken from the experimental findings obtained on male subjects, i.e., gap acceptance rates for short gaps in condition of frustration were 0.37 if the oncoming traffic was manually operated and 0.97 if it was an HAV; likewise, detection rate of frustration was 0.78.

In the uncontrolled setting, SiSAT based on this MDP computed the risk of traversing though a too short gap as being in the interval [0.96999999, 0.97000001] for the worst possible scenario. I.e., SiSAT managed to construct the worst-case scenario where the human driver gets frustrated when an HAV is upcoming and the gap in front of it is too short, provoking a risk that coincides with the short-gap acceptance likelihood of 0.97 against HAVs. With the aforementioned control strategy from Section III.E, the worst-case risk of traversing though a too short gap was computed as [0.29584999, 0.29585001], implying a risk reduction by a factor of approx. 3.3 despite the uncertainties in sensing frustration by neurophysiological measurements. While these findings cannot directly be transferred to realistic driving situations due to lack of a sufficiently dense empirical basis of some model elements, they certainly are indicative for the relevance of the approach.

## IV. CONCLUSION AND FUTURE WORK

Based on the example of using functional near-infrared spectroscopy for detecting frustration of drivers in oncoming traffic and adapting an HAV's driving strategy accordingly, we exposed an architecture employing neuro-physiological sensors and human driver models for optimizing availability of HAVs without impeding their safety. The quantitative risk analysis proves that the inherent uncertainty in measuring and interpreting human cognitive states is no show-stopper in safety-critical environments, thus clearly demonstrating the commercial value proposition of the approach.

REFERENCES

[1] Steven E. Shladover, Presentation given at Workshop on Societal and Technological Research Challenges for Highly Automated Road Transportation Systems in Germany and the US, October 30, 2018, Washington D.C.
[2] Unni, A., Ihme, K., Jipp, M., & Rieger, J. W. (2017). Assessing the driver's current level of working memory load with high density functional near-infrared spectroscopy: a realistic driving simulator study. *Frontiers in human neuroscience*, *11*, 167.
[3] Ihme, K., Unni, A., Zhang, M., Rieger, J. W., & Jipp, M. (2018). Recognizing Frustration of Drivers from Face Video Recordings and

Brain Activation Measurements with Functional Near-Infrared Spectroscopy. *Frontiers in human neuroscience*, *12*.

[4] Villringer, A., Planck, J., Hock, C., Schleinkofer, L., & Dirnagl, U. (1993). Near infrared spectroscopy (NIRS): a new tool to study hemodynamic changes during activation of brain function in human adults. *Neuroscience letters*, *154*(1-2), 101-104.

[5] Solovey, E. T., Zec, M., Garcia Perez, E. A., Reimer, B., & Mehler, B. (2014). Classifying driver workload using physiological and driving performance data. In *Proceedings of the 32nd annual ACM conference on Human factors in computing systems - CHI '14* (pp. 4057–4066). New York, New York, USA: ACM Press. https://doi.org/10.1145/2556288.2557068

[6] Gable, T. M., Kun, A. L., Walker, B. N., & Winton, R. J. (2015). Comparing heart rate and pupil size as objective measures of workload in the driving context. In *Adjunct Proceedings of the 7th International Conference on Automotive User Interfaces and Interactive Vehicular Applications - AutomotiveUI '15* (pp. 20–25). New York, New York, USA: ACM Press. https://doi.org/10.1145/2809730.2809745

[7] Sander, D., Grandjean, D., & Scherer, K. R. (2005). A systems approach to appraisal mechanisms in emotion. *Neural Networks*, *18*(4), 317–352. https://doi.org/10.1016/j.neunet.2005.03.00

[8] De Waard, D. (1996). *The Measurement of Drivers' Mental Workload*. Groningen University, Netherlands. Retrieved from http://apps.usd.edu/coglab/schieber/pdf/deWaard-Thesis.pdf

[9] Parasuraman, R., & Rizzo, M. (2006). *Neuroergonomics*. Oxford University Press. https://doi.org/10.1093/acprof:oso/9780195177619.001.0001

[10] Young, M. S., Brookhuis, K. A., Wickens, C. D., & Hancock, P. A. (2015). State of science: mental workload in ergonomics. *Ergonomics*, *58*(1), 1–17

[11] Scherer, K. R. (2005). What are emotions? And how can they be measured? *Social Science Information*, *44*(4), 695–729. https://doi.org/10.1177/0539018405058216

[12] Kramer, A. F., & Parasuraman, R. (2007). Neuroergonomics: Applications of Neuroscience to Human Factors. In J. T. Cacioppo, L. G. Tassinary, & G. Berntson (Eds.), *Handbook of Psychophysiology* (pp. 704–722). Cambridge: Cambridge University Press. https://doi.org/10.1017/CBO9780511546396.030

[13] Villringer, A., Planck, J., Hock, C., Schleinkofer, L., & Dirnagl, U. (1993). Near infrared spectroscopy (NIRS): a new tool to study hemodynamic changes during activation of brain function in human adults. *Neuroscience Letters*, *154*(1–2), 101–4. https://doi.org/10.1016/0304-3940(93)90181-J

[14] Villringer, A., & Chance, B. (1997). Non-invasive optical spectroscopy and imaging of human brain function. *Trends in Neurosciences*, *20*(10), 435–42. Retrieved from http://www.ncbi.nlm.nih.gov/pubmed/9347608

[15] Hastie, T., Tibshirani, R., & Friedman, J. H. (2009). *The elements of statistical learning : data mining, inference, and prediction* (2nd ed.). Springer. Retrieved from http://www.springer.com/de/book/9780387848570

[16] Qian, J., Hastie, T., Friedman, J., Tibshirani, R., & Simon, N. (2013). Glmnet for Matlab 2013. *URL Http://www. Stanford. Edu/~ Hastie/glmnet_matlab*.

[17] Trende, A., Unni, A., Weber L., Rieger, J. W., Luedtke, A., (2019, June). "Differences in behavior in human-human and human-autonomous vehicle interactions in time-critical merging situations". Submitted to PETRA '18 the 12th Pervasive Technologies Related to Assistive Environments Conference, Rhodes, Greece – June 5-7, ACM, New York, NY, USA

[18] Ebbesen, E. B., & Haney, M. (1973). Flirting with Death: Variables Affecting Risk Taking at Intersections 1. *Journal of Applied Social Psychology*, *3*(4), 303-324.

[19] Barber, D. (2012). *Bayesian reasoning and machine learning*. Cambridge University Press.

[20] Sivak, M., & Schoettle, B. (2012). A note: The changing gender demographics of US drivers. *Traffic injury prevention*, *13*(6), 575-576.

[21] Toledo, T., Koutsopoulos, H., & Ben-Akiva, M. (2003). Modeling integrated lane-changing behavior. *Transportation Research Record: Journal of the Transportation Research Board*, (1857), 30-38.

[22] Hwang, S. Y., & Park, C. H. (2005). Modeling of the gap acceptance behavior at a merging section of urban freeway. In *Proceedings of the Eastern Asia Society for Transportation Studies* (Vol. 5, No. 1641, p. e1656). Tokyo: Eastern Asia Society for Transportation (EASTS).

[23] Davis, G., & Swenson, T. (2004). Field study of gap acceptance by left-turning drivers. *Transportation Research Record: Journal of the Transportation Research Board*, (1899), 71-75.

[24] Pollatschek, M. A., Polus, A., & Livneh, M. (2002). A decision model for gap acceptance and capacity at intersections. *Transportation Research Part B: Methodological*, *36*(7), 649-663.

[25] Farah, H., Polus, A., Bekhor, S., & Toledo, T. (2007). Study of passing gap acceptance behavior using a driving simulator. *Advances in Transportation Studies an International Journal*, 9-16.

[26] Fränzle, M., Teige, T., & Eggers, A. (2010). Engineering constraint solvers for automatic analysis of probabilistic hybrid automata. *The Journal of Logic and Algebraic Programming*, *79*(7), 436-466.

[27] Chen, M., Fränzle, M., Li, Y., Mosaad, P. N., & Zhan, N. What's to come is still unsure: synthesizing controllers resilient to delayed interaction. In *ATVA 2018*: 56-74. LNCS 11138, Springer Verlag 2018.

[28] Minderhoud, M. M., & Bovy, P. H. (2001). Extended time-to-collision measures for road traffic safety assessment. *Accident Analysis & Prevention*, *33*(1), 89-97.

[29] Salvucci, D. D. (2006). Modeling driver behavior in a cognitive architecture. *Human factors*, *48*(2), 362-380.

[30] Lüdtke, A., Weber, L., Osterloh, J. P., & Wortelen, B. (2009, July). Modeling pilot and driver behavior for human error simulation. In *International Conference on Digital Human Modeling* (pp. 403-412). Springer, Berlin, Heidelberg.

[31] Weber, L., Baumann, M., Lüdtke, A., & Steenken, R. (2009). Modellierung von Entscheidungen beim Einfädeln auf die Autobahn. *To appear: Fortschritts-Berichte VDI: Der Mensch im Mittelpunkt technischer Systeme*, *8*.

[32] Wortelen, B., Unni, A., Rieger, J. W., & Lüdtke, A. (2016, October). Towards the integration and evaluation of online workload measures in a cognitive architecture. In *Cognitive Infocommunications (CogInfoCom), 2016 7th IEEE International Conference on* (pp. 000011-000016). IEEE.

[33] Damm, W., & Galbas, R. (2018, May). Exploiting learning and scenario-based specification languages for the verification and validation of highly automated driving. In *2018 IEEE/ACM 1st International Workshop on Software Engineering for AI in Autonomous Systems (SEFAIAS)* (pp. 39-46). IEEE.